# Digital Radio-over-Multicore-Fiber System with Self-Homodyne Coherent Detection and Entropy Coding for Mobile Fronthaul

Lu Zhang[1,2], Aleksejs Udalcovs[3], Rui Lin[1], Oskars Ozolins[3], Xiaodan Pang[1,3,6], Lin Gan[4], Richard Schatz[1], Anders Djupsjöbacka[3], Jonas Mårtensson[3], Ming Tang[4], Songnian Fu[4], Deming Liu[4], Weijun Tong[5], Sergei Popov[1], Gunnar Jacobsen[3], Weisheng Hu[2], Shilin Xiao*[2], Jiajia Chen*[1]

[1] KTH Royal Institute of Technology, Kista, Sweden, jiajiac@kth.se
[2] State Key Laboratory of Advanced Optical Communication System and Networks, Shanghai Jiao Tong University, Shanghai, China, slxiao@sjtu.edu.cn
[3] NETLAB – Networking and Transmission Laboratory, RISE AB, Kista, Sweden
[4] Huazhong University of Science and Technology, Wuhan, China
[5] Yangtze Optical Fiber and Cable Joint Stock Limited Company, Wuhan, China
[6] Infinera, Fredsborgsgatan 24, 117 43 Stockholm, Sweden

**Abstract** *We experimentally demonstrate a 28-Gbaud 16-QAM self-homodyne digital radio-over-33.6km-7-core-fiber system with entropy coding for mobile fronthaul, achieving error-free carrier aggregation of 330 100-MHz 4096-QAM 5G-new-radio channels and 921 100-MHz QPSK 5G-new-radio channels with CPRI-equivalent data rate up to 3.73-Tbit/s.*

## Introduction

To tackle with the continuous increase of capacity demand in mobile fronthaul (MFH), improving system spectrum efficiency and system capacity is of great importance. Although analog radio-over-fiber (RoF) system has high spectrum efficiency, its nonlinear distortions significantly affect the transmission quality and distance, making it hardly meet the requirements for 5G new-radio (NR)[1], particularly when high modulation orders (e.g. 1024-QAM) are needed. As a result, digital RoF system is more favored for MFH. Common public radio interface (CPRI) protocol[2] is standardized to transmit digitized analog signal between distributed units (DUs) and remote units (RUs) for MFH, e.g., 4G-MFH and 5G-MFH-I Option-8[2]. CPRI requires high capacity due to a large number of quantization bits (QBs). To address the capacity crunch, there are mainly two ways: 1) increasing the system capacity by implementing coherent transmission[3]; 2) reducing the number of QBs using compression coding[4-6]. However, 5G-NR, where full dimension massive-MIMO is expected to implement active-phased-array-antennas with the array size of 128 and possibly beyond, could make the existing CPRI based solutions not sufficient to meet the MFH capacity requirement anymore.

In this paper, we report our recent demonstration of digital radio-over-multicore-fiber system for MFH, capable of CPRI-equivalent data rate beyond 1-Tbit/s. The demonstrated system utilizes coherent detection and multi-core-fiber (MCF) to increase the MFH capacity, where the local oscillator (LO) signal uses one core for self-homodyne coherent detection. Meanwhile, to reduce the number of QBs, two entropy coding (EC) schemes, namely Huffman coding (HC) and arithmetic coding (AC), are introduced. Combining differential coding[5] and EC techniques for MFH, we experimentally demonstrate a 28-Gbaud 16-QAM self-homodyne digital radio-over-33.6km-7-core-fiber system for MFH, realizing error-free aggregation of 330 100-MHz 4096-QAM 5G-NR carriers and 921 100-MHz QPSK 5G-NR carriers with CPR-equivalent data rate up to 3.73-Tbit/s.

## Entropy coding (EC) schemes

As the air interface of 5G-NR, orthogonal frequency division multiplexing (OFDM) signals always show high peak-to-average-power ratio. In digital MFH, most of the quantized signals, i.e., codewords, need to be distributed in the low-amplitude region and very few codewords are required in the high-amplitude region. It causes entropy redundancy, since all codewords share the same number of QBs. EC is promising to reduce such entropy redundancy. The idea of EC is very simple[7], where the codewords occurred more frequently are implemented by shorter codes while the codewords occurred less frequently are implemented by longer codes. We introduce both HC (see Fig. 1) and AC (see Fig. 2) to digital MFH.

For HC, the codewords are sorted by their probabilities of occurrences in descending order. Assuming in total $M$ number of QBs, there are $2^M$ codewords $\{A_1, A_2, \ldots A_{2^M}, p(A_1) > p(A_2) > \ldots > p(A_{2^M})\}$, where $p(A_i)$ represents the probability that codeword $A_i$ occurs. Fig. 1 shows how to construct a Huffman code-tree. Each codeword is represented by a node with weight as its probability of occurrence. Two nodes with the lowest probabilities are chosen and assigned with value **0** and **1,** respectively. Then these two nodes are merged into a new node with the weight as sum of weight of its two child nodes. If there is still unmerged node, the algorithm continues to find two nodes with the lowest weight, which are assigned value **0** and **1,** respectively, and then merged to a new node. Finally, for every codeword (i.e., a leaf node in the Huffman code-tree), the coded bits are read from the root node to the leaf node. For HC, the average codeword length $l_c$ is bounded by:

$$E\{A\} \le l_c \cdot 2^M < E\{A\}+1, \quad (1)$$

where $E\{A\} = \sum_{i=1}^{2^M} p(A_i)\log_2 p(A_i)$ is the entropy of the codeword.

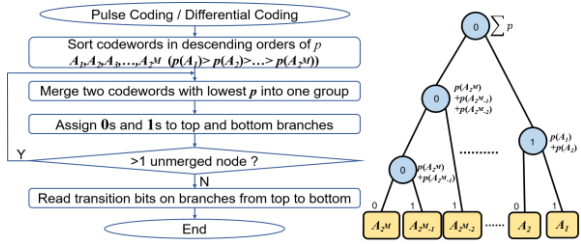

**Fig. 1:** Flowchart of HC and the typical code-tree

Although HC approaches the entropy of quantization, the code length for each codeword should be integer, which produces rounding errors[7]. AC overcomes this problem in HC. Besides, in AC the coding process is carried out at arrival of each codeword, which does not need a 'priori' to determine the mappings of all codewords in advance. Fig. 2 shows how to construct arithmetic code-tree. In the initial stage, all possible codewords are in the range [$u_0$, $v_0$), where $u_0$=0.0, and $v_0$=1.0. In every iteration, a new quantization word $x_n=A_i$ with a cumulative probability range [$low_i$, $high_i$] is handled, where $low_i = \sum_{j=1}^{i-1} p(A_j)$ and $high_i = \sum_{j=1}^{i} p(A_j)$, and the range [$u_n$, $v_n$) is updated accordingly, where $u_n = u_{n-1}+(v_{n-1}-u_{n-1})*low_i$ and $v_n = u_{n-1}+(v_{n-1}-u_{n-1})*high_i$. Finally, when all quantization words are handled, a probability $\tilde{p} = (u_n+v_n)/2$ is selected. For example, if $\tilde{p} = 0.6640625 = 0.1010101_2$, and the entire codeword is **1010101**. For AC, the average codeword length $l_c$ is also bounded according to Eq.1. However, the compression efficiency of AC is always better or identical to that of HC[7].

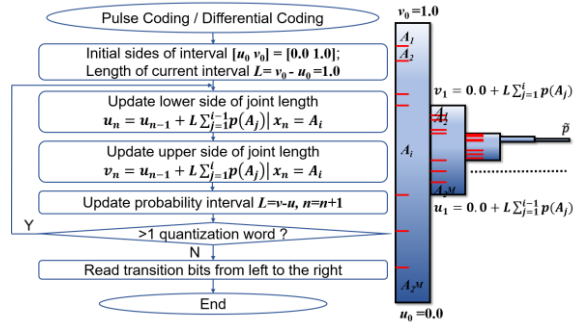

**Fig. 2:** Flowchart of AC and the typical code-tree

### Experimental setup and results

The experimental setup is shown in Fig. 3. Self-homodyne coherent detection scheme with MCF is adopted for experimental demonstrations. OFDM and 2048-inverse fast Fourier transfer (IFFT) points are used. The sampling rate per OFDM symbol is 122.88-MSa/s with the bandwidth of 100-MHz. At the transmitter side, the quantization sequence is generated by either pulse-coding or differential-coding and then coded by either HC or AC. They are mapped to 16-QAM constellations followed by Nyquist pulse shaping. The 16-QAM signals are output from the AWG (50-GSa/s) and then loaded to the IQ-modulator. The light from the semiconductor laser is split using a PM 50:50 splitter. One is used as the transmitter laser, and the other one transmitted after MCF is used as the LO at the receiver side. The output light from the IQ modulator is amplified by an EDFA, divided into 6 branches using a 1:8 splitter, decorrelated, and then coupled into the low-crosstalk 7-core MCF via a fan-in. After 33.6-km MCF, the channels are demultiplexed by a fan-out. A noise loading module consisting of a VOA and an EDFA, is placed to adjust a channel optical signal-to-noise ratio (OSNR). Finally, the signals are coherently detected and stored in the DSO (80-GSa/s). Carrier phase recovery is used for coherent equalization. The bit-error ratio (BER) curve is shown in Fig. 3(a). For all cores, we have achieved error-free 16-QAM transmissions.

The probability distributions of the quantization codewords after pulse coding (i.e., CPRI) and differential coding[5] are shown in Fig.

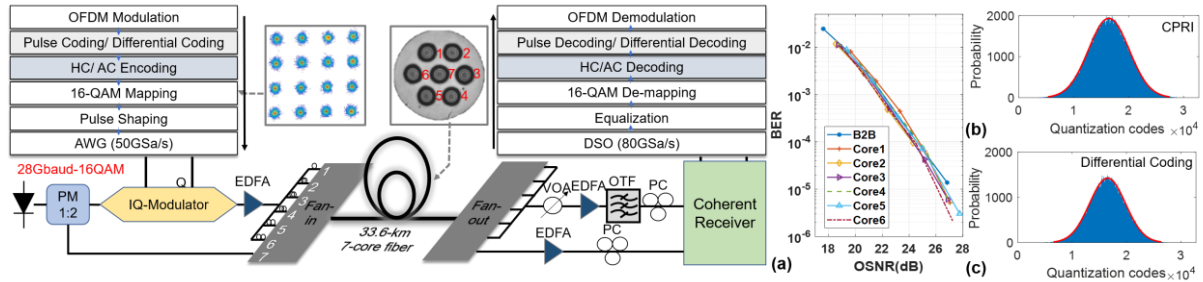

**Fig. 3:** Experimental setup, (a) BER curve of the 16-QAM self-homodyne MFH, (b-c) probability distribution of quantization codewords of CPRI and differential coding (15QBs). AWG: arbitrary waveform generator, VOA: variable optical attenuator, EDFA: Erbium doped fiber amplifier, PM: polarization maintaining, DSO: digital storage oscilloscope

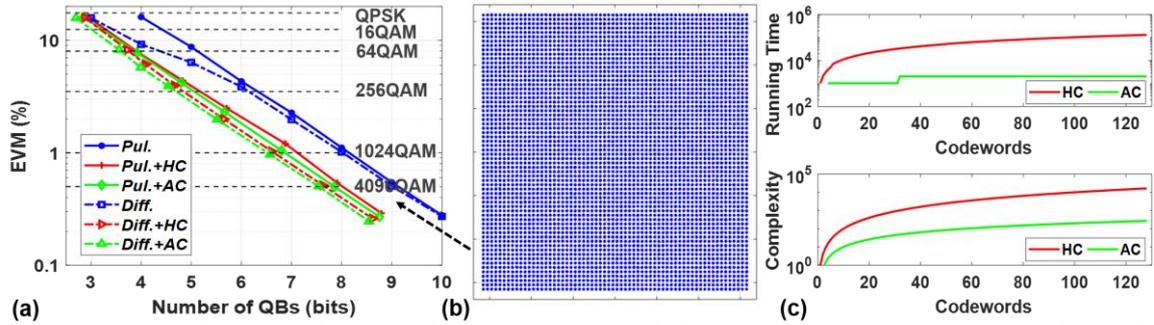

**Fig. 4:** EVM versus QBs of (a) PCM and DPCM, (b) recovered 4096QAM, (c) running time and complexity of HC and AC

3(b) and Fig. 3(c), respectively. The quantization codewords fit the Gaussian distribution and the codewords in the low-amplitude region occur more frequently. The experimental error-vector-magnitude (EVM) performance for 5G-NR with different modulation orders is shown in Fig. 4(a). They are measured with error-free coherent transmissions over 33.6-km MCF. First, we measure the EVM performance with pulse code modulation (PCM) and EC, where the number of QBs after EC is calculated as effective QBs (the number of QB before EC/compression-ratio). In average, HC reduces ~1.3 QBs and AC reduces ~1.5 QBs. For CPRI (i.e. PCM with 15 QBs), the number of QBs can be reduced to 13.4275 after HC while the number of QBs can be reduced to 13.3983 after AC. Moreover, we measure the EVM performance with differential pulse code modulation (DPCM)[5] and EC. Our previous work[5] has proved that differential coding outperforms pulse coding for MFH, particularly at low number of QBs, as it reduces the quantization noise by adaptive linear prediction. With EC, the performance of differential coding based MFH can be further improved. At low number of QBs, there is less entropy redundancy of the differential coding, and the reduction of QBs is less than 1 after EC. ~1.4 QBs can be reduced after HC and ~1.6 QBs can be reduced after AC at high number of QBs. Finally, with only 7.5385 QBs, 4096-QAM based 5G-NR signals can be delivered over MFH with error-free transmission[8]. The recovered 4096-QAM constellation for 5G-NR is shown in Fig. 4(b) with EVM at ~0.5%.

It can be observed that AC outperforms HC in terms of QBs (saving ~0.2 QBs). Besides, AC also outperforms HC in terms of running time and computational complexity, which are shown in Fig. 4(c). However, the floating-point precision may influence AC, so that AC needs to be processed in short blocks to avoid performance deterioration.

The experimental aggregated 5G-NR channels and CPRI-equivalent data rates using differential coding with and without EC are concluded in Table.1. The number of aggregated channels and equivalent CPRI rates are calculated as follows [a], where 16/15 MIMO overhead and 66b/64b line coding are considered.

**Table. 1:** Experimental aggregated 5G-NR channels and the equivalent CPRI rates

| QAM order | Aggregated 5G-NR channels | | | Equivalent CPRI rates (Tbit/s) | | |
|---|---|---|---|---|---|---|
| | Diff. | Diff.+HC | Diff.+AC | Diff. | Diff.+HC | Diff.+AC |
| 4 | 829 | 864 | 921 | 3.36 | 3.5 | 3.73 |
| 16 | 621 | 668 | 698 | 2.52 | 2.71 | 2.83 |
| 64 | 497 | 608 | 622 | 2.02 | 2.47 | 2.52 |
| 256 | 355 | 440 | 451 | 1.44 | 1.78 | 1.83 |
| 1024 | 311 | 376 | 378 | 1.26 | 1.53 | 1.53 |
| 4096 | 276 | 326 | 330 | 1.12 | 1.32 | 1.34 |

$Chs = 28*4*6/\{0.12288*2*QB*(16/15)*(66/64)\}$, (2)
$CPRI\text{-}equivalent\ data\ rate = 28*4*6*15/QB$. (3)

**Conclusions**

With self-homodyne coherent transmission, MCF and effective coding, MFH system is experimentally demonstrated to support the highest modulation order up to 4096 QAM, with 330 100-MHz 5G-NR channels leading to CPRI-equivalent data rate of 1.34-Tbit/s, and the highest CPRI-equivalent data rate of 3.73-Tbit/s with carrier aggregation of 921 100-MHz QPSK 5G-NR channels.

**Acknowledgements**

We wish to thank the Swedish Research Council (VR), the Swedish Foundation for Strategic Research (SSF), Göran Gustafsson Foundation, the Swedish ICT-TNG, EU H2020 MCSA-IF Project NEWMAN (#752826), VINNOVA funded SENDATE-EXTEND and SENDATE-FICUS, National Natural Science Foundation of China (#61331010, 61722108, 61775137, 61671212).

---

[a] In real implementation, error-coding schemes may be introduced to guarantee the error-free MFH transmission, where the results in Table. 1 will be reduced by the its overhead (e.g. 7% for HD-FEC).